\documentstyle[twocolumn]{mn}

\def\deg{\ifmmode^\circ\else$^\circ$\fi}

\def\kpc{\ifmmode h^{-1}{\rm kpc}\else$h^{-1}{\rm kpc}$\fi}
\def\kms{\ifmmode {\rm km~s}^{-1}\else${\rm km~s}^{-1}$\fi}
\def\bii{\ifmmode b^II\else b$^{II}$\fi}
\def\lii{\ifmmode l^II\else l$^{II}$\fi}
\def\feh{\ifmmode {\rm [Fe/H]}\else [Fe/H]\fi}

\title[Metallicities for Edinburgh-Cape stars]
{Metallicity estimates for A-, F-,  and G-type stars from the
Edinburgh-Cape blue object survey}
\author[T. C. Beers et al.]
       {T. C. Beers,$^1$ S. Rossi,$^2$ D. O' Donoghue,$^3$ D. Kilkenny,$^3$ R.
S. Stobie,$^3$ C. Koen,$^3$
\newauthor and R. Wilhelm$^4$\\
$^1$Department of Physics \& Astronomy, Michigan State University, E.
Lansing, MI  48824 \\
$^2$Instituto Astron\^omico e Geof\'isico, Universidade de S\~ao Paulo,
Av. Miguel Stefano 4200, 04301-904, S\~ao Paulo, Brazil\\
$^3$South African Astronomical Observatory, PO Box 9, Observatory 7935,
South Africa\\
$^4$McDonald Observatory, University of Texas, Austin, TX  78712}

\date{Accepted 2000 June 30.
      Received 2000 June 29;
      in original form 2000 June 28}

\pagerange{\pageref{firstpage}--\pageref{lastpage}}
\pubyear{2000}

\begin{document}

\maketitle

\label{firstpage}

\begin{abstract}
The Edinburgh-Cape Blue Object Survey is an ongoing project to identify and
analyse a large sample of hot stars selected initially on the basis of
photographic colours (down to a magnitude limit $B \sim 18.0$) over the entire
high-Galactic-latitude southern sky, then studied with broadband $UBV$
photometry and medium-resolution spectroscopy.  Due to unavoidable errors in
the initial candidate selection, stars that are likely metal-deficient dwarfs
and giants of the halo and thick-disk populations are inadvertently included,
yet are of interest in their own right.  In this paper we discuss a total of
206 candidate metal-deficient dwarfs, subgiants, giants, and horizontal-branch
stars with photoelectric colours redder than $(B-V)_0 = 0.3 $, and with
available spectroscopy.  Radial velocities, accurate to $\sim 10-15$ \kms , are
presented for all of these stars.  Spectroscopic metallicity estimates for
these stars are obtained using a recently re-calibrated relation between Ca
{\sc II} K-line strength and $(B-V)_0$ colour.  The identification of
metal-poor stars from this colour-selection technique is remarkably efficient,
and competitive with previous survey methods.  An additional sample of 186 EC
stars with photoelectric colours in the range $-0.4 \le (B-V)_0 < 0.3$,
comprised primarily of field horizontal-branch stars and other, higher-gravity,
A- and B-type stars, is also analysed.  Estimates of the physical parameters
T$_{\rm eff}$, log g, and \feh\ are obtained for cooler members of this
subsample, and a number of candidate RR Lyrae variables are identified.

\end{abstract}


\section{INTRODUCTION}

Deep, and in particular, {\it complete} surveys of stars in the halo of the
Galaxy always turn up objects of astrophysical interest outside of the original
targets of the survey.  The Edinburgh-Cape Blue Object Survey (hereafter, EC
survey) is no different in this respect (see Stobie et al. 1997 for a detailed
description).  Briefly, the EC survey is intended to discover large numbers of
stars in the southern sky with colours normally associated with the blue
stellar populations of the Galaxy, such as white dwarfs and subdwarf O- and
B-type stars, in a similar fashion to the Palomar-Green survey in the northern
hemisphere (Green, Schmidt, \& Liebert 1986).  The initial colour-selection 
criterion for EC stars makes use of a $B$,$U-B$ polygon in the (photographic)
magnitude-colour diagram (see Fig. 2 of Stobie et al. 1997).  The attempt was
made, on a plate-by-plate basis, to set the red boundaries of this polygon in
such a way as to exclude most of the F- and G-type stars, while retaining most
of the bluer objects.  The shape of the boundary differs from plate to plate,
and naturally becomes somewhat less clear at fainter magnitudes, due to the
increasing errors in the photographic photometry.  For the EC survey, a
contamination level of $\sim $20 per cent in the original selection was deemed
acceptable.  During the course of spectroscopic and photometric follow-up
observations of the EC stars brighter than $B \sim 16.5$, it has become clear
that many of these `contaminant' stars are in fact of great interest as probes
of the metal-poor stellar populations of the Galaxy.  The present paper
explores the nature of these stars.

In this paper we report radial velocities, stellar classifications, and
spectroscopic abundance estimates, expressed in terms of \feh\ , for 321
main-sequence turnoff (TO), subgiant (SG) , giant (G), and field
horizontal-branch (FHB) or other A-type stars identified in the process of the
EC survey follow-up campaign.  Radial velocities (only) are reported for 73
stars which are likely B-type horizontal-branch (HBB) stars.  We note that the
present sample contains 108 stars with radial velocity and abundance
determinations in the Galactic anti-rotation direction at lower Galactic
latitudes ($|b| \le 45^\circ $); these data provide valuable information
concerning the rotation of the halo and thick-disk components of the Galaxy
(see, e.g., Chiba \& Beers 2000).

\section{OBSERVATIONS AND DATA REDUCTION}

Broadband $UBV$ measurements of EC survey stars have been previously
obtained with photoelectric photometers on the SAAO 0.75m and 1.0m telescopes.
The majority of the stars in the present paper have had apparent magnitudes and
colours reported in Kilkenny et al. (1997) -- these have been supplemented with
some newer data.  Column (1) of Table 1 lists the EC number of the stars in the
present sample.  The 37 stars in this sample that exhibit possible
peculiarities of one kind or another are identified with superscripts attached
to their names.  Columns (2) and (3) list the equinox 1950.0 coordinates for
each star.  Errors in these positions are typically $\pm 2$ arcsec.  Columns
(4) and (5) are the Galactic longitude and latitude, respectively.  Columns
(6)-(8) report the $UBV$ photometry.  Stobie et al. (1997) show that the
typical errors in the EC photoelectric photometry vary from 0.02 to 0.04
mag as one progresses from the brighter ($V < 13.5 $) to the faintest
stars ($V > 16.5$). Reddening estimates in the direction toward each star are
obtained by interpolation of the Burstein \& Heiles (1982) maps.  An initial
reddening estimate is made assuming the star under consideration lies
completely above or below the Galactic dust layer (taken here to have a scale
height of $h = 125$ pc).  The reddening to a given star at distance D (where
available) is reduced compared to the total reddening by a factor
$[1-\exp(-|D\; \sin\; b|/h)]$, where $b$ is the Galactic latitude.  This
estimate, which accuracy typically not better than 0.03 mag, is listed
in column (9).  For stars that we are unable to assign distances to, we simply
adopt the unaltered reddening estimate from Burstein \& Heiles.

Medium-resolution ($\sim$ 3.5 \AA\ over two pixels) spectra of EC survey stars
were obtained with the RPCS spectrograph (using an intensified Reticon photon
counting detector + grating \# 6) on the SAAO 1.9m telescope, over the spectral
range  3600 \AA\ $\le \lambda \le 5200$ \AA\ , and with a typical
signal-to-noise $10/1 \le {\rm S/N} \le 15/1$.  Data reduction was performed
using the SAAO program {\sc SKIP}; a detailed description of the procedures can
be found in Stobie et al. (1997).

F- and G-type stars are readily recognized from the spectra based on the
strengths of the Ca {\sc II} H and K lines, centered on $\lambda \sim 3950$
\AA\ , and the CH G-band feature at $\lambda \sim 4300$ \AA\ .  Others objects
among our program stars appear to be FHB- or A-type stars, based on the depth
and breadth of their Balmer lines and the occasional presence of HeI absorption
features.  Sample spectra of the stars analysed in this paper are shown in Fig.
9 of Stobie et al. (1997).

Fig. 1a shows a histogram of the distribution of apparent $V$ magnitudes for
our program stars.  Roughly 60 per cent of the EC stars in the present sample
are fainter than about $V = 14.5$, which, for main sequence luminosities or
greater, puts them at distances farther than 750 pc away, well outside the thin
disk, and approaching the scale height of the thick disk (in the vertical
directions).  If follows that this sample may contain a substantial fraction of
metal-deficient stars. Fig. 1b shows the distribution of de-reddened $(B-V)_0$
colours.  It is essentially bi-modal in appearance, with the great majority of
the bluer stars being identified below as likely HBB stars.

Fig. 2 is the de-reddened two-colour diagram for the program stars, shown with
a solar-abundance main-sequence line from Johnson (1966) superposed.  One can
clearly recognize dwarfs of low metal abundance that, due to the de-blanketing
effect, lie above the main-sequence solar-metallicity line .  In addition,
there are a number of metal-poor FHB stars that lie close to this line due to
the effect of low surface gravity on their colours.  A handful of stars with
colours we would associate with the Blue Metal-Poor (BMP) stars of Preston,
Beers, \& Shectman (1994) are present as well, with $0.0 \le (B-V)_0
\le 0.35$ and $-0.5 \le (U-B)_0 \le -0.1$.  The grouping of stars with colours
in the range $-0.4 \le (B-V)_0 \le 0.0$ and $-1.0 \le (U-B)_0 \le -0.1$ are
dominated by horizontal-branch B-type (HBB) stars, but may additionally include
main-sequence and subdwarf B-type stars.

A number of stars in our program sample have been identified by previous
surveys.  Table 2 lists these identifications, obtained by conducting 
a search of the {\sc SIMBAD} database for stars with coordinates (and
commensurate colours or spectral types) within 1 arcmin of the positions listed
in Table 1.

\section{DERIVED PARAMETERS}

\subsection{Radial velocities and line indices}

Radial velocities are obtained for each of our program stars using the
line-by-line and cross-correlation techniques, described in detail in Beers et
al. (1999), and references therein.  Application of these techniques to the
slightly higher-resolution spectra of the HK survey follow-up effort (1-2 \AA\
) yielded velocities accurate to on the order of 7--10 \kms\ , so we expect
that our present results should only be slightly worse than this, on the order
of $\sim 10-15 $ \kms.  This level of accuracy is still sufficient to be useful
for analysis of the kinematics for the thick-disk and halo populations, since
their velocity dispersions are in the range $\sim 45 - 150 $ \kms\ .
Measurements of the heliocentric radial velocities, after correction for the
Earth's rotation and orbital motion, are listed in column (10) of Table 1.

For each star, the derived (geocentric) radial velocities are used to place a
set of fixed bands for the derivation of line-strength indices, which are
pseudo-equivalent widths of prominent spectral features.  The bands we employ
are summarized in Table 3.  A complete discussion of the choice of bands, and
the `band-switching' scheme used to produce our derived Ca {\sc II} K-line
index, KP, and Balmer-line index, HP2, is provided in Beers et al. (1999).

Line indices (in \AA\ ) for each of the program stars are reported in columns
(11) -- (14) of Table 1.  Column (12) lists He {\sc I} line indices (at
wavelengths $\lambda = 4026$ \AA\  and $\lambda = 4471 $ \AA\ ), where
available, with the blue and the red line listed in the left and right sides of
the column, respectively.  In order for a line-index measurement to be
considered a detection, we require that the derived indices reach a minimum
value of 0.25 \AA\ .  The He {\sc I} indices (and the G-band index GP) are
useful in the detection of likely `composite stars' --  stars that exhibit a
combination of spectra 0and colours that indicate the possible presence of a
companion star of different spectral type than the classification we have
adopted (see Fig. 3 of Kilkenny et al. 1997 for other examples from the EC
survey).

\subsection{Stellar classifications}

Stellar classifications, based on visual inspection of the spectra, and
consideration of the observed $UBV$ photometry, are provided in column (15)  of
Table 1.  For the majority of the program stars, these classifications are the
same as those given in Kilkenny et al. (1997).  However, a number of stars
in the present sample did not appear in the previously published list, hence
classifications for these stars are shown here for the first time.

For the purpose of further analysis it is convenient to divide the sample
according to colour.  We first consider the subsample of stars satisfying $-0.4
\le (B-V)_0 \le 0.4$, and refer to these stars below as the `hot-star'
subsample.  Stars with colours  $(B-V)_0 \ge 0.3$ are considered `cool stars.'
The overlap of these two samples in the range $0.3 \le (B-V)_0 \le 0.4$ is used
to provide a check on the abundance determinations obtained from two
different calibrations.  Although the calibrations were constrained by
comparison with very different stellar samples, the two calibrations do share a
common model-atmosphere source, hence they are not totally independent of one
another.

\subsection{The horizontal-branch B-type stars}

There are 73 stars among our program sample that are likely to be hot
horizontal-branch B-type (HBB) stars; these are listed in Table 4.  For stars
at such high temperatures it is not feasible to obtain estimates
of physical parameters based on analysis of medium-resolution spectra.  It is
possible that some of the stars listed in Table 4 are {\it not} HBB stars at
all (alternatives would include sdB, as well as post-AGB stars -- see
discussions in Hambly et al. 1997 and Kendall et al. 1997), but we have no
means of making further refinements in their classifications based on the
present data.   A discussion of their radial velocity distribution is presented
in \S 4 below.

There are four stars in Table 4 that fall in the colour range $0.0 \le (B-V)_0
\le 0.4$ and $ (U-B)_0 \le 0.4$ that may be composite stars -- these are
identified in the table with a colon (:) following their names.

\subsection{Physical parameter estimates for the hot stars}

Wilhelm, Beers, \& Gray (1999a) discuss the development and calibration of a
spectroscopic and photometric technique which enables the identification and
classification of FHB- and other A-type stars over the temperature range $6000
\le {\rm T}_{\rm eff} \le 10000$ K.  This technique makes use of broadband
$UBV$ colours predicted from model atmosphere calculations, Balmer-line
profiles, Ca {\sc II} K equivalent widths, and a synthetic-template comparison
method, to estimate the physical parameters T$_{\rm eff}$, log g, and
\feh\ , with precision on the order of $\sigma ({\rm T}_{\rm eff}) = \pm
250$~K, $\sigma (\log {\rm g}) = \pm 0.25$ dex, and $\sigma (\feh ) = \pm 0.3$
dex, respectively.  Wilhelm et al. (1999b) apply this methodology to a large
sample of hot stars identified in the HK objective-prism survey, and discuss
some of the advantages and pitfalls of this approach.  As part of their
high-resolution follow-up study of BMPs, Preston \& Sneden (2000) have
confirmed that the abundances for BMPs derived by Wilhelm et al. (1999b) are
well-matched with those obtained from their own analysis, giving additional
confidence in the estimates of metallicity obtained by these methods.  In \S
3.7 below, we discuss an `internal' comparison of abundance determination for
stars in the overlapping colour regions of the hot and cool subsamples.

Table 5 summarizes the results obtained by application of the Wilhelm et al.
(1999a) technique to the hot-star subsample.  Column (1) lists the EC star
name.  Columns (2) and (3) list estimates of effective temperature and surface
gravity, respectively.  Note that for stars with colours at or near the blue
limits of the Wilhelm et al. grids of model atmospheres ($(B-V)_0 = -0.25$
and/or $(U-B)_0 = -0.45)$, the estimated temperatures and surface gravities may
be subject to larger errors.  These stars are indicated with an appended colon
following the estimated parameter.

Columns (4) and (5) of Table 5 list estimates of metal abundance obtained from
analysis of the Ca {\sc II} K line.  The metallicity estimate $\feh_{{\rm
W}_{\rm K}}$ is based on the measured equivalent width of the Ca {\sc II} K
line.  The estimate $\feh_{\rm CTA}$ is obtained from profile fits to this same
line.  For the lower signal-to-noise spectra of some program stars, these
estimates occasionally differ markedly -- this is especially true for the
hotter stars with very weak Ca {\sc II} K lines.  Column (6) lists estimates of
abundance, $\feh_{\rm MTA}$, obtained by comparison of two `metallic-line
regions,' falling in the wavelength intervals 4175 \AA\ $ \le \lambda \le 4310
$ \AA\ and 4360 \AA\ $ \le \lambda \le 4500 $ \AA\ ,  with template synthetic
spectra.  Following the precepts of Wilhelm et al., we adopt an abundance
estimate $\feh_{\rm AVG}$, listed in column (7), that is the mean of the two
most-consistent estimators from columns (4)-(6).  Column (8) lists the absolute
difference in the abundance determinations from the most-consistent estimators
from one another.  It should be kept in mind that the lowest abundance that can
be reliably obtained for stars in the hot-star subsample is $\feh = -3.0$.
Again, in the case of stars with colours at or near the grid limits of the
models, a colon is appended to their estimated abundances to indicate
additional uncertainty.

Column (9) of Table 5 lists the stellar type classifications obtained from
application of the Wilhelm et al. methods.  Four classes are used:  FHB -- a
likely field-horizontal branch star; A -- a star with main-sequence gravity
typical of A-type stars, BMPs, and other, slightly cooler, early F-type stars;
FHB/A -- an indeterminate classification, and Am -- a star with apparently
main-sequence gravity and a large discrepancy between the metallic-line-region
abundance estimate and those estimates based on the Ca {\sc II} K line, a
phenomenon often associated with Am (and Ap) stars (see Wilhelm et al. 1999a,
1999b for more details).  For the stars in Table 5 classified as Am, the
appropriate abundance estimate is that listed in column (6), $\feh_{\rm MTA}$.

Column (12) of Table 5 identifies a number of stars with the entry `r' that
exhibit discrepancies between their Balmer-line strengths and their observed
broadband colours, which might arise from spectroscopic and photometric
observations of an RR Lyrae variable taken `out of phase.'  These
identifications are tentative, and should be confirmed on the basis of further
photometric follow-up.  Note that several stars in this table are known RR
Lyrae variables (indicated with an `R' in column (12)).

There are 21 stars in Table 5 that are identified in Table 1 as having possible
peculiarities in their spectra and/or colours, leading to somewhat less
confidence in the proper assignment of their spectral classifications and in
the derived estimates of their physical parameters.  These stars are indicated
in Table 5 with an appended colon following their adopted class in column (9).
A colon has also been appended to their physical parameter estimates listed in
the table.

\subsection{Distance estimates for the hot stars}

For the stars in Table 5 classified as FHB, estimates of absolute magnitude and
distance (in parsecs) are derived following the procedure described by Wilhelm
et al. (1999b), and are listed in Columns (10) and (11), respectively.  For
stars classified as A or Am, we have adopted an estimate of distance,
calibrated by the main sequences of young open clusters (shifted appropriately
to account for their metal abundances), as described by Beers et al. (2000).
This methodology differs from that used by Wilhelm et al., and produces
distance estimates for these stars that are as much as a factor of two
{\it less} than those authors report for stars of the same classification (note
that the previous procedure did not explicitly take metallicity into account,
producing much of the discrepancy).  No estimated distances are given for the
stars classified as FHB/A.

\subsection{Abundance determinations for the cool stars}

Beers et al. (1999) describe a technique for the estimation of [Fe/H] from
medium-resolution spectroscopy of stars, based on the strength of the Ca {\sc
II} K line as a function of de-reddened $(B-V)_0$ colour, with accuracy on the
level of 0.15--0.2 dex over the abundance range $-4.0 \le \feh \le 0.0$.  The
EC spectra are not generally of sufficiently high signal-to-noise to make use
of the Auto-Correlation Function technique described by Beers et al., however,
we can still make use of their correction matrices to ensure that,
in particular, the more metal-rich cool stars do not have their abundances
significantly {\it under}-estimated by the Ca {\sc II} K-line technique.

We have employed this methodology to obtain abundance estimates for the 206
cool EC stars with available spectra and colours redder than $(B-V)_0 = 0.3$.
The results of the abundance analysis are summarized in Table 6.  Column (1)
lists the EC star.  Column (2) lists the type classification obtained from the
precepts described in Beers et al.  (1999), with the caveat described below.
The Beers et al. classification scheme is based on a comparison of the derived
de-reddened colour with the expected location of a star in the colour-magnitude
diagram  for stars of a variety of ages (from 5 to 15 Gyrs) and metallicities
($-3.0 \le {\rm [Fe/H]} \le 0.0$) obtained from the Revised Yale Isochrones
(Green 1988; King, Demarque, \& Green 1988).  Column (3) lists the estimated
absolute magnitude, $M_V$, and column (4) lists the associated distance
estimate (in parsecs).  In cases where the type classification is ambiguous,
e.g., TO/FHB or FHB/TO, alternative absolute magnitudes and distances are
provided in parentheses.  Column (5) lists the estimated metallicity obtained
by application of the Beers et al.  calibration, [Fe/H]$_{\rm K3}$, and its
associated one-sigma error.  For the 12 cool stars noted in Table 1 with
peculiarities, we have attached a colon to their type classifications and
metallicity estimates.

The caveat in the type classifications mentioned above concerns our separation
of FHB and TO stars in the colour range $0.3 \le (B-V)_0 \le 0.5$, which
differs from that used previously in the HK survey follow-up.   The original
classification scheme (dating back to Beers, Preston, \& Shectman 1985) was
designed for application to stars of very low metallicity ([Fe/H] $< -2.0$).
However, since we are now obtaining abundance estimates for stars with
metallicities up to the solar value, the procedure by which the `split' between
FHB and TO stars has to be modified.  In the original scheme, stars in the
colour range $0.3 \le (B-V)_0 \le 0.5$, which includes both FHB and TO stars,
were assigned classifications based on the simple criteria:

\begin{eqnarray*}
 TO:\;\; & \; & (U-B)_0 \le -0.1 \\ [0.25in]
FHB:\;\; & \; & (U-B)_0 > -0.1 
\end{eqnarray*}

\noindent This split is not adequate for stars with abundances [Fe/H] $> -2.0$.

Fig. 3 shows the `limiting' $(U-B)_0$ colours as a function of metallicity, as
predicted from the Revised Yale Isochrones for (a) stars with surface gravities
$\log\; g \le 3.0$ (appropriate for FHB stars in this colour range--see Wilhelm
et al. 1999a) and (b) stars with $\log\; g \ge 4.0$ (appropriate for TO stars
in this colour range).  For the present classification exercise, we make the
FHB/TO assignments by comparison with these limiting lines (broadened slightly
to include estimated reddening errors in $(U-B)_0$ on the order of 0.02
mag).  Stars with $(U-B)_0$ colours {\it above} the TO line (i.e., at
lower values of $(U-B)_0$) are classified as TO, while those {\it below} the
FHB line (i.e., at higher values of $(U-B)_0$) are classified as FHB.  Stars of
intermediate surface gravities fall between the lines, and their
classifications are less certain, but we assign a class based on the closest of
the two limiting lines.  In most cases accurate photometry should produce a
reliable assignment.  For stars with ambiguous classifications, we have
attached a colon to the abundance determination listed in Table 6 to indicate
the additional uncertainty in this estimate.

\subsection{Comparison of abundance determinations}

There are 26 stars in the colour region $0.3 \le (B-V)_0 \le 0.4$ that appear
in both the hot- and cool-star subsamples discussed above, and hence
have metallicity estimates determined from two different calibrations.  Fig.
4 shows a comparison of these estimates.  The agreement between the two
metallicity estimates is generally good, with a slight tendency for the more
metal-deficient stars determined from the hot-star calibration to have derived
abundances that are higher than those determined from the cool-star
calibration.  The difference between these estimates, in the sense $\feh_{\rm
AVG} - \feh_{\rm K3}$, as quantified by the biweight estimate of central
location discussed by Beers, Flynn, \& Gebhardt (1990), is $C_{BI}= +0.1$ dex.
The biweight scale of their difference is $S_{BI} = 0.5$ dex, only slightly
higher than the expected error obtained from the square root of the quadratic
sum of the errors of the methods, approximately 0.4 dex.

\section{RADIAL VELOCITY DISTRIBUTIONS FOR THE EC STARS}

Fig. 5 displays histograms of the heliocentric radial velocities for the
three subsamples of EC stars considered in this paper.  The distributions are
quite similar to one another, and all exhibit dispersions consistent with stars
selected from the Galactic halo population, $\sigma_{\rm los} \sim 100-120$
\kms\ .  The appearance of the cool-star subsample distribution suggests the
presence of a low-dispersion population, which is likely to include members of
the metal-weak thick disk of the Galaxy -- note that Chiba \& Beers (2000) have
shown the metal-weak thick disk extends to metallicities as low as $\feh =
-2.0$.  The hot-star distribution appears to have little or no evidence for the
presence of this low-dispersion population -- a result of the generally larger
distances explored by this subsample as compared to the cool-star subsample.
Fig. 6 shows a comparison of the cumulative distributions of these two
subsamples as a function of height above the Galactic plane.  Over half ($\sim
65$ per cent) of the cool-star subsample is located within 1 kpc of the plane, the
region most likely to be dominated by metal-weak thick-disk stars, while only
$\sim$25 per cent of the hot-star subsample is located this close to the plane.

Fig. 7 displays the distribution of heliocentric radial velocities for the
cool-star and hot-star subsamples as a function of estimated \feh\ .  The
high-velocity stars from these subsamples are clearly drawn from similar
(likely, identical) parent populations, while the presence of a
low-dispersion population of metal-weak stars down to $\feh \sim -2.0$ is clear
among the cool-star subsample.

\section{EFFICIENCY OF THE EC SURVEY FOR DETECTION OF VERY METAL-POOR STARS}

Figs. 8a and 8b show the distribution of derived metal abundances for the
hot-star and cool-star subsamples, respectively.  Figs. 9a and 9b are the
cumulative distributions for these same data sets, which we compare below to
the results of other surveys for metal-poor stars.

A variety of techniques have been used previously to discover metal-poor stars
in the Galaxy.  The approaches can be broadly divided into two categories --
proper-motion-selected stars, such as those discussed by Ryan \& Norris (1991)
and Carney et al. (1994), and objective-prism surveys, such as the HK survey
described by Beers, Preston, \& Shectman (1992).  Inspection of the cumulative
metallicity distribution of the EC stars in the cool-star subsample, shown in
Fig. 9a, suggests that the discovery efficiency of stars with abundances less
than $\feh = -1.0$ is significantly better than that obtained from
spectroscopic follow-up of proper-motion-selected stars shown in Fig. 9c.  The
cumulative distribution function of metallicity for the HK-survey stars, shown
in Fig. 9d, appears to be superior for the identification of large numbers of
stars with $\feh \le -1.0$, but since it was targeted at such stars this is
perhaps not a great surprise.  Applying the definition of the `effective yield'
(EY) of survey techniques for extremely metal-poor stars described by Beers
(2000) (numbers of stars discovered with $\feh \le -2.0$ compared to the number
of stars inspected), the relevant numbers for the samples shown in Figs. 8a-d
are: EY (EC cool stars) = 0.19, EY (EC hot stars) = 0.18, EY
(proper-motion-selected stars) = 0.11, EY (objective-prism-selected stars) =
0.50.

We wish to emphasize that the EYs of the EC stars quoted above is made based on
the {\it entire} set of stars examined in this paper.  Clearly, if our purpose
was simply to select the most likely metal-poor candidates, the $UBV$
photometry obtained during the course of the EC follow-up would allow us to
substantially increase the EY of this technique.  For instance, if we were to
restrict the colours to only include stars in the range $0.3 \le (B-V)_0 \le
0.5$, the EY of the EC survey would rise dramatically, to on the order of 0.30.
If one further restricted the stars of interest in this colour range to include
only stars with $(U-B)_0 \le -0.1$, the EY of the EC survey would rise to 0.40.
This result bodes well for future colour-selected samples of metal-poor
candidates, such as those obtained from the Sloan Digital Sky Survey (e.g.,
Lenz et al.  1993).  Christlieb \& Beers (2000) describe a refined
selection technique based on automated classification of prism spectra from
the Hamburg/ESO Objective-Prism Survey that achieves EY = 0.80 for
metal-deficient dwarfs near the halo main-sequence turnoff {\it without} the
need for a photometric pre-filter.

One might wonder whether the relatively high EY of the EC survey could result
because, at faint apparent magnitudes, one is exploring deep enough into the
halo population of the Galaxy that a random sample of stars might be expected
to contain a large fraction of metal-deficient objects.  That is, how much has
the (original, photographic) colour selection actually helped with the
identification of truly metal-poor stars ?   One way to empirically address
this question is by making a comparison of the metallicity distribution
function (MDF) of the cool stars obtained at different distances above or below
the Galactic plane.  Figs. 10a and 10b are comparisons of the low-abundance
tail of the MDF for the subset of stars with $ |Z| \le 1000$ pc with that for
the subset of stars with $ |Z| > 1000$ pc.  Only the low-abundance tails have
been compared to avoid complications introduced by the presence of the
metal-weak thick disk.  As can be seen in the Figures, although there does
exist a slight tendency for the more distant stars to include objects of
greater metal deficiency, it is not a particularly strong one, and may be
influenced by the small number statistics.  Furthermore, Reid \& Majewski
(1993) have shown that in order to obtain a `pure' sample of halo stars
(without contamination from the tail of the substantially more populous
thick-disk component) one must reach out to distances greater than $\sim 5-7$
kpc above or below the Galactic plane.  For stars near the main-sequence
turnoff, which dominate the EC cool-star subsample, this would correspond to
apparent magnitudes as faint as $V \sim 18.5$, substantially fainter than the
limiting magnitude of the stars for which photometry and spectroscopy have been
obtained during the EC survey follow-up.

Of particular interest are the 36 stars in the cool-star subsample with $\feh
\le -2.0$; the lowest-metallicity cool star has an abundance $\feh = -2.94$.
The hot-star subsample includes another 20 stars with $\feh
\le -2.0$.  Stars of such low metallicity are quite rare in the Galaxy, at
least brighter than the apparent magnitude limits that have been explored to
date.

We expect that spectroscopic observations of additional stars from the
EC survey that satisfy the colour criteria described herein, but which
at present do not have spectroscopic data, will result in the discovery of
additional stars with $\feh \le -2.0$.  Based on the results found to date, as
the EC survey progresses, we might expect the discovery of another $\sim 250$
stars with $\feh \le -2.0$, for a total of $\sim 300$ stars with abundance less
than 1 per cent of solar, and a total of $\sim 1250$ stars with $\feh \le -1.0$.  By
way of comparison, the total yield of stars with $\feh \le -1.0$ in the HK
survey follow-up to date numbers $\sim 2000$ stars.  Hence, a dedicated
campaign to obtain medium-resolution spectroscopy of the EC `contaminants' can
make a substantial contribution to the database of known metal-deficient stars
in the Galaxy.

\section* {ACKNOWLEDGEMENTS}

TCB acknowledges support for this work from grant AST 95-29454 from the
National Science Foundation.  SR acknowledges partial support for this
work from grant 200068/95-4 CNPq, Brazil, and from the Brazilian Agency FAPESP.
We also thank a semi-anonymous referee for comments and suggestions which
improved the presentation.

This work made use of the SIMBAD database, operated at CDS, Strasbourg, France.

\newpage

\section* {FIGURE CAPTIONS}

\noindent{\bf Figure 1.} (a) Distribution of photoelectric $V$ magnitudes for
the EC survey stars considered herein. (b) Distribution of photoelectric
de-reddened $(B-V)_0$ colours for the EC survey stars.
\bigskip

\noindent{\bf Figure 2.} De-reddened two-colour diagram for the EC survey
stars.  The line shown is a solar-abundance main-sequence gravity line taken
from Johnson (1966). The crosses correspond to stars noted in Table 1 with some
peculiarities (binary, possible variable, composite) associated with them.
Errors in the photometry (on the order of 0.02 mag) are too small to be
usefully shown on the scale of this plot.

\bigskip

\noindent{\bf Figure 3.} Limiting $(U-B)_0$ colours of stars in the colour
range $0.3 \le (B-V)_0 \le 0.5$, as a function of metallicity, obtained from
the Revised Yale Isochrones.  The region above the solid line corresponds to
the locations of stars with surface gravities typical of main-sequence TO
stars, while the region below the dashed line applies for stars with surface
gravities of FHB stars.  See text for additional explanation.
\bigskip

\noindent{\bf Figure 4.} Comparison of abundances estimated for EC survey stars
in the colour range $0.3 \le (B-V)_0 \le 0.4$ determined using the techniques
of Wilhelm et al.  (1999a), $\feh_{\rm AVG}$, with those determined using the
techniques of Beers et al. (1999), $\feh_{\rm K3}$.  The solid line is the
one-to-one line.  The dashed line is a locally weighted regression line.  The
agreement between the two techniques is good down to $\feh \sim -2.0$.  Below
this abundance, the Wilhelm et al. technique appears to slightly overestimate
the derived abundance as compared to the cool-star calibration. 
\bigskip

\noindent{\bf Figure 5.} Radial velocities for the three subsamples of EC
survey stars -- (a) the EC cool stars, (b) the EC hot stars, and (c) the EC HBB
stars.  In each case, the solid line indicates the best-fit (single) Gaussian
distribution.
\bigskip

\noindent{\bf Figure 6.} Cumulative distribution functions for the EC cool-star
(filled circles) and hot-star (open circles) subsamples, respectively, as a
function of absolute distance from the Galactic plane.  Note that the cool-star
subsample contains a large fraction of stars with distances less than 1 kpc
from the plane, while only a minor fraction of the hot-star subsample lies this
close.
\bigskip

\noindent{\bf Figure 7.} Radial velocities as a function of estimated
metallicity for the EC cool stars (filled circles) and the EC hot stars (open
circles).  Note that both distributions are quite similar.  The cool stars, in
addition to a halo component, appear to include a low velocity-dispersion
population of stars over the metallicity interval $-2.0 \le \feh \le 0.0$, most
likely associated with the metal-weak thick disk. 
\bigskip

\noindent{\bf Figure 8.} Histograms of the estimated abundances for (a) the EC
hot stars, and (b) the EC cool stars.
\bigskip

\noindent{\bf Figure 9.} Cumulative distribution functions of metallicity for
stars selected by various survey techniques -- (a) the EC cool stars, (b) the
EC hot stars, (c) high-proper-motion stars from Carney et al. (1994), and (d)
HK survey stars, selected by the objective-prism technique. 
\bigskip

\noindent{\bf Figure 10.} Comparison of the metallicity distribution functions
for two subsets of the cool-star subsample with $\feh \le -2.0$. (a) A
histogram of the two distributions. The dashed histogram is for stars with
distances in excess of 1 kpc from the plane, while the solid histogram is for
stars closer than 1 kpc.  Note that, with the exception of a single bin at low
abundance, the histograms are quite similar to one another.  (b) Cumulative
distribution functions of the two distributions.  The open circles connected by
the dashed line are for those stars with distances in excess of 1 kpc from the
plane, while the filled circles connected by the solid line are for those stars
closer than 1 kpc.  Note that the cumulative distributions are quite similar to
one another. 

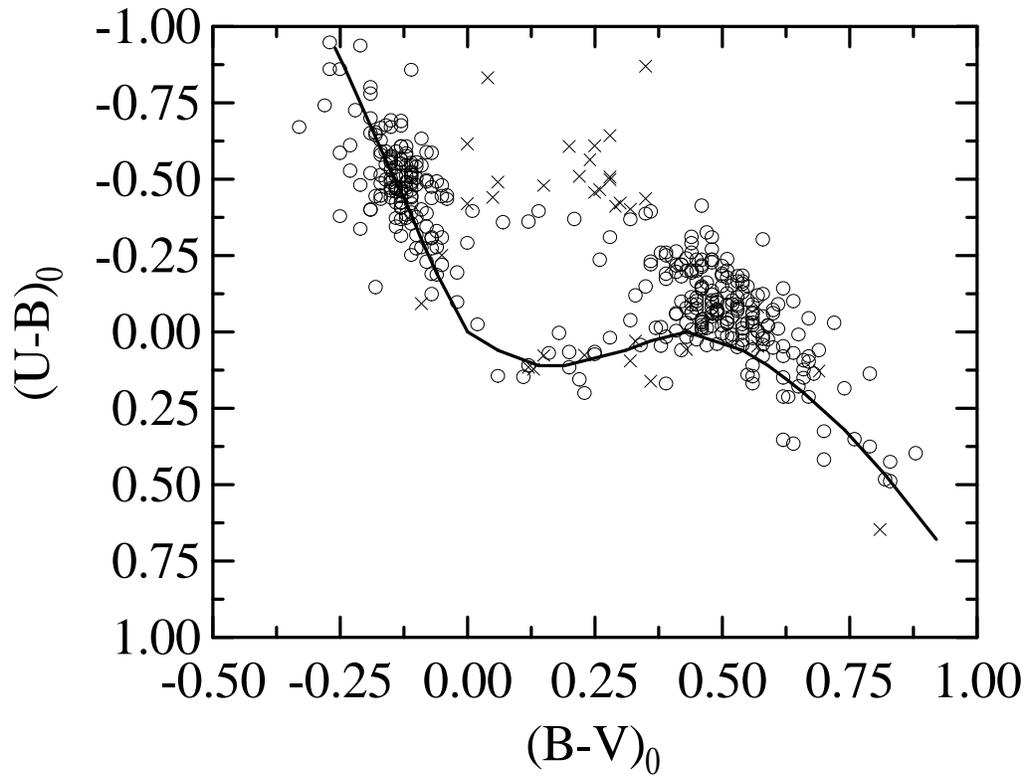
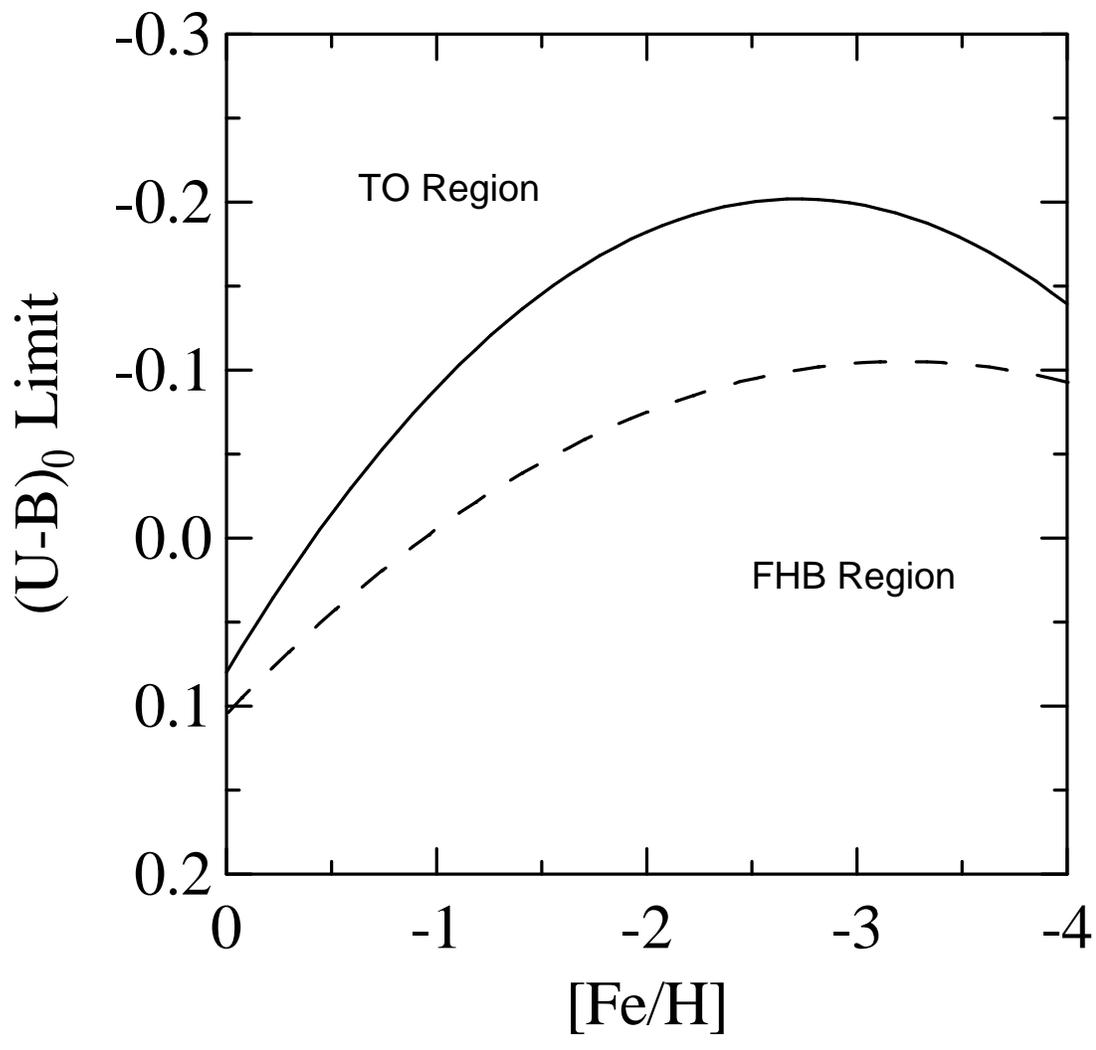
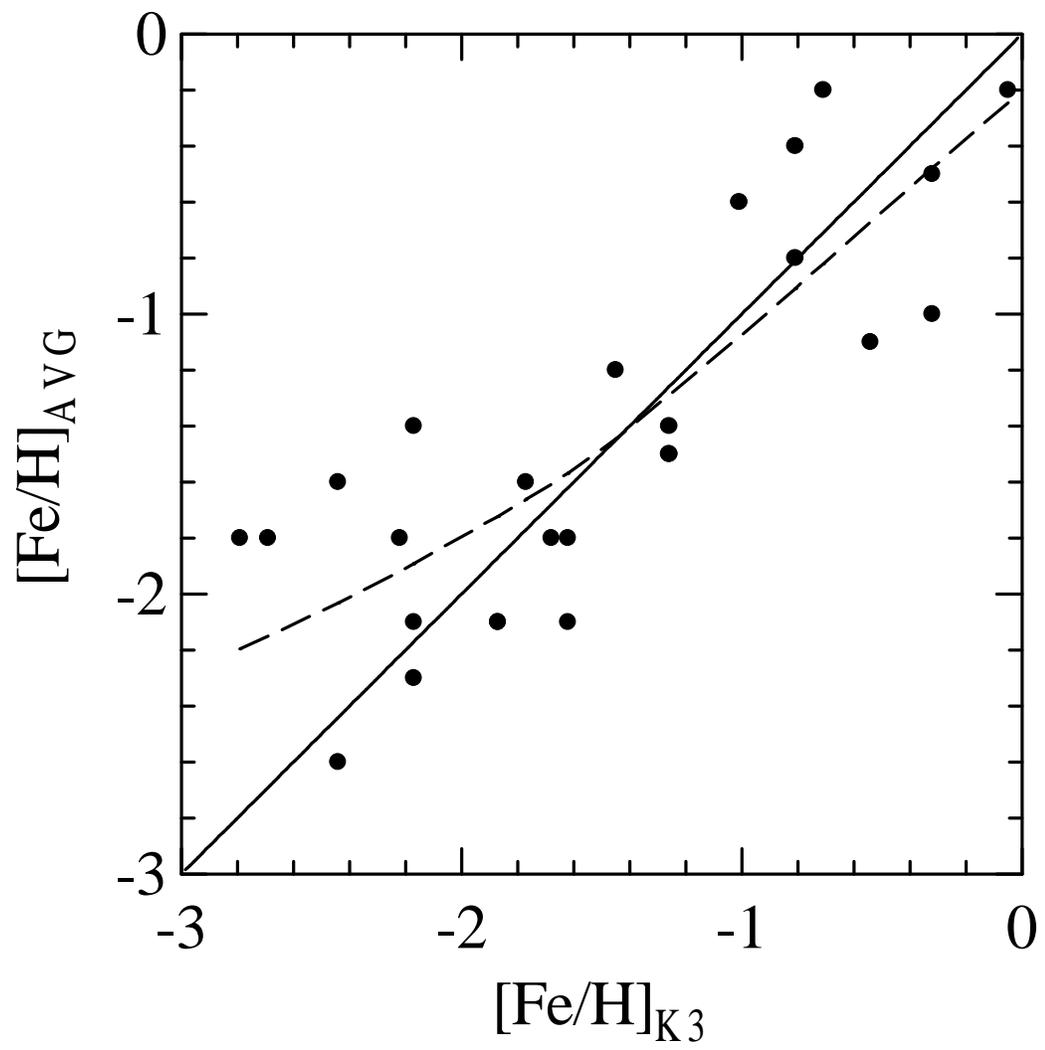
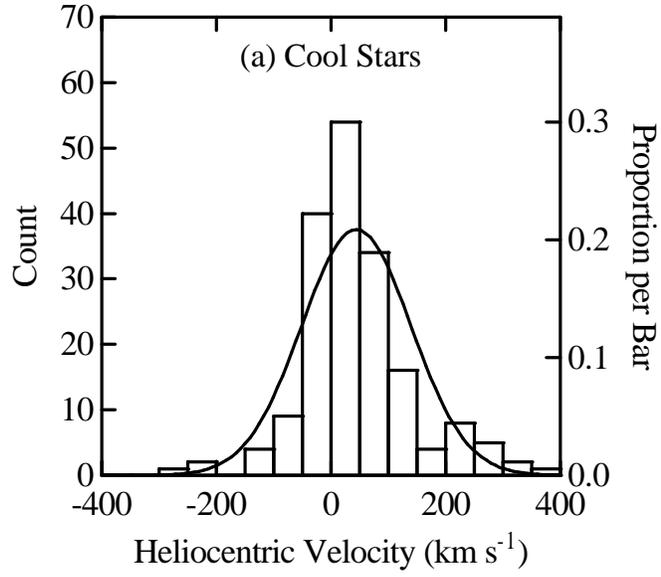
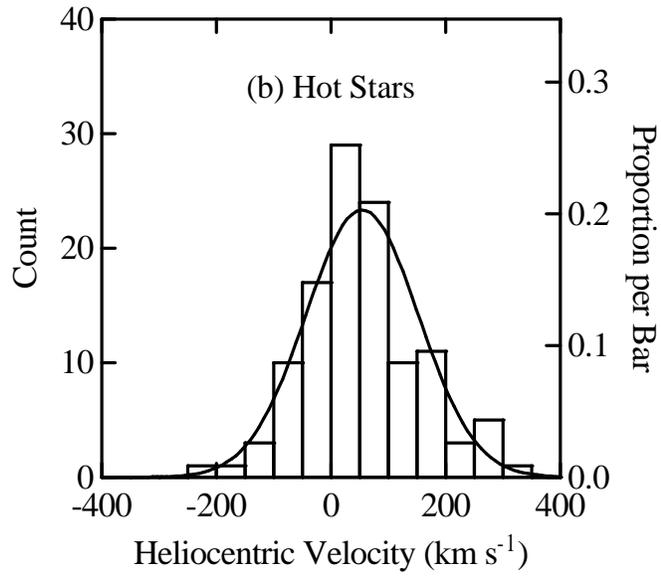
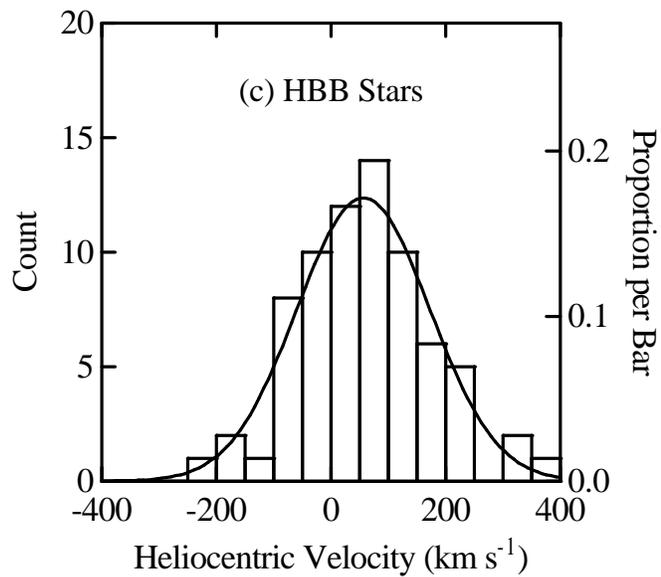
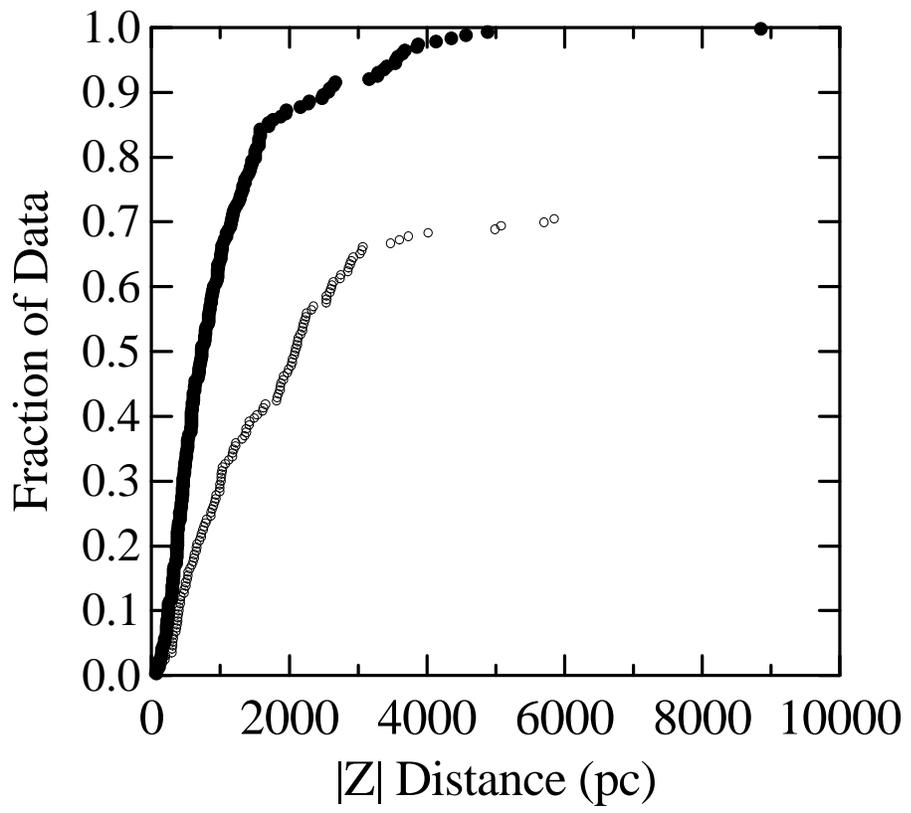
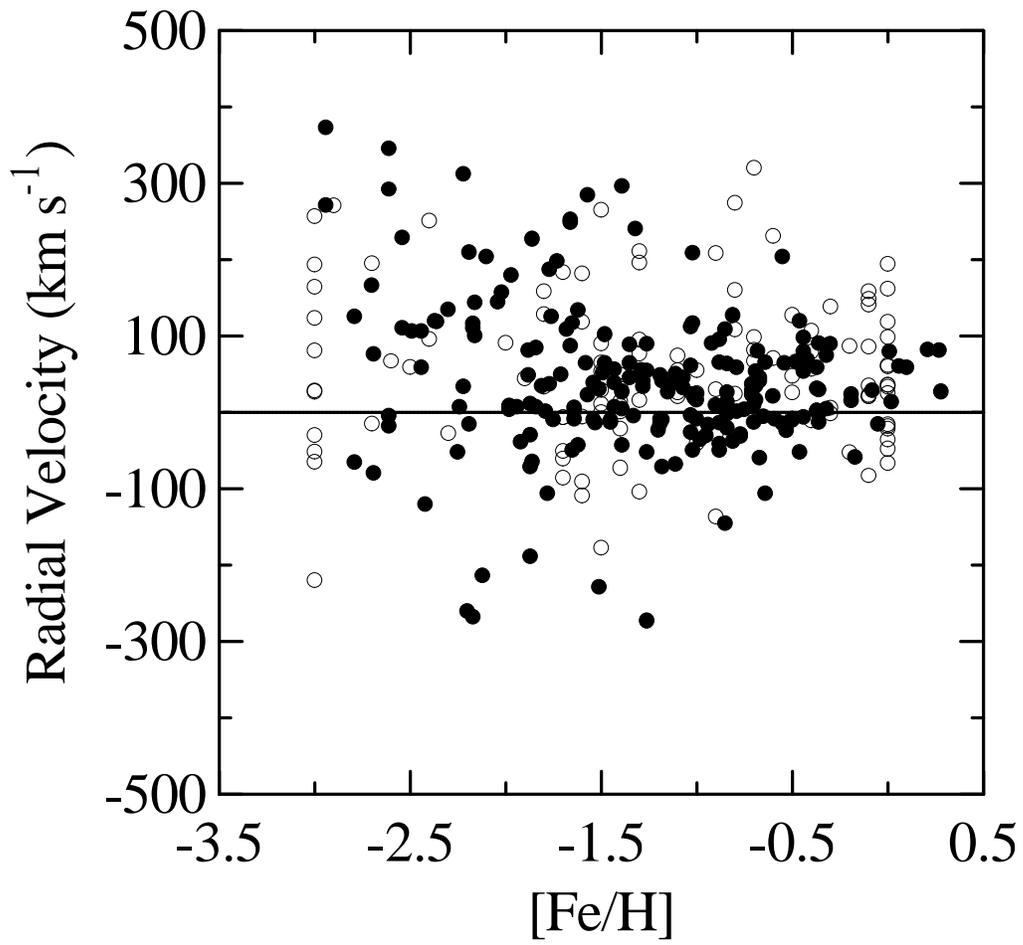
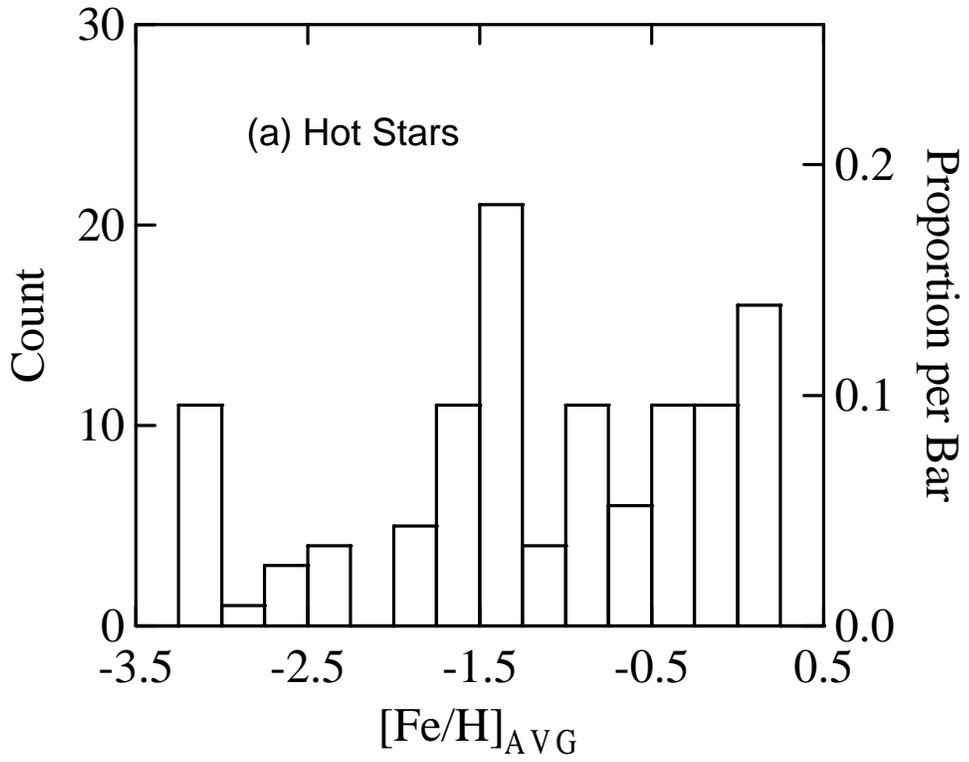
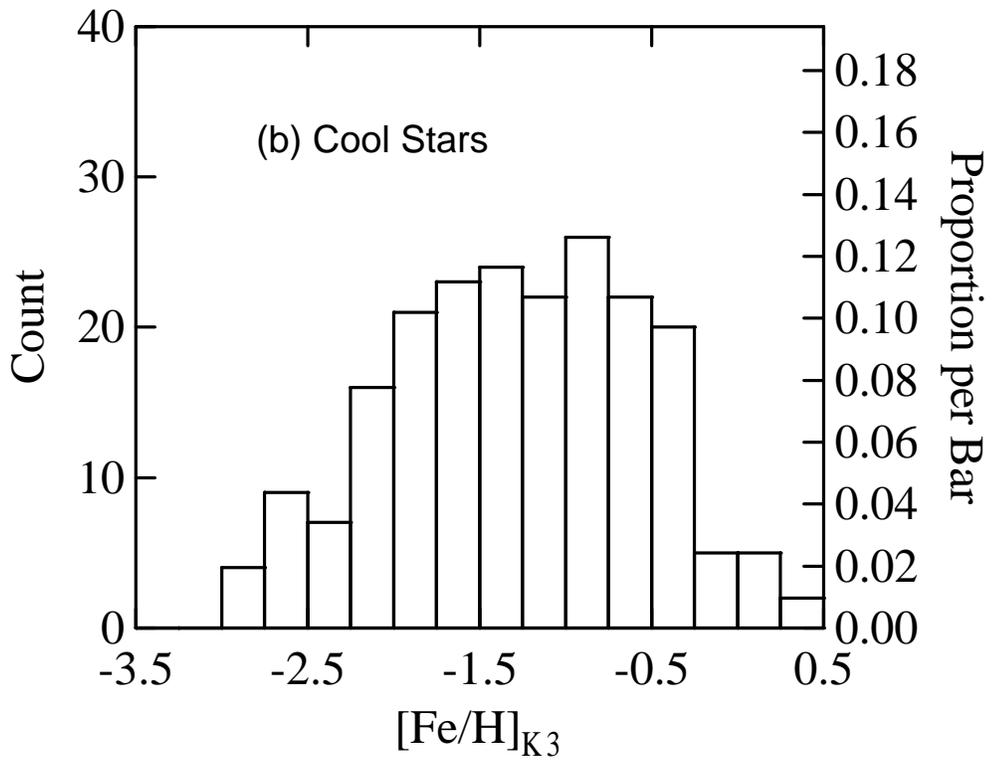
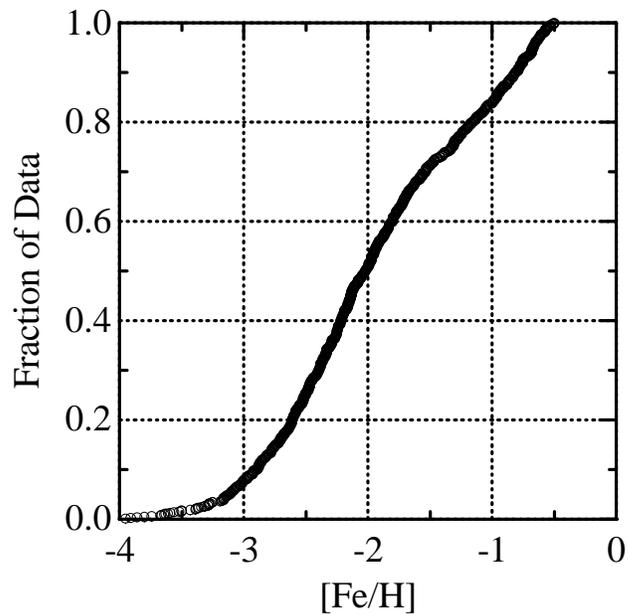
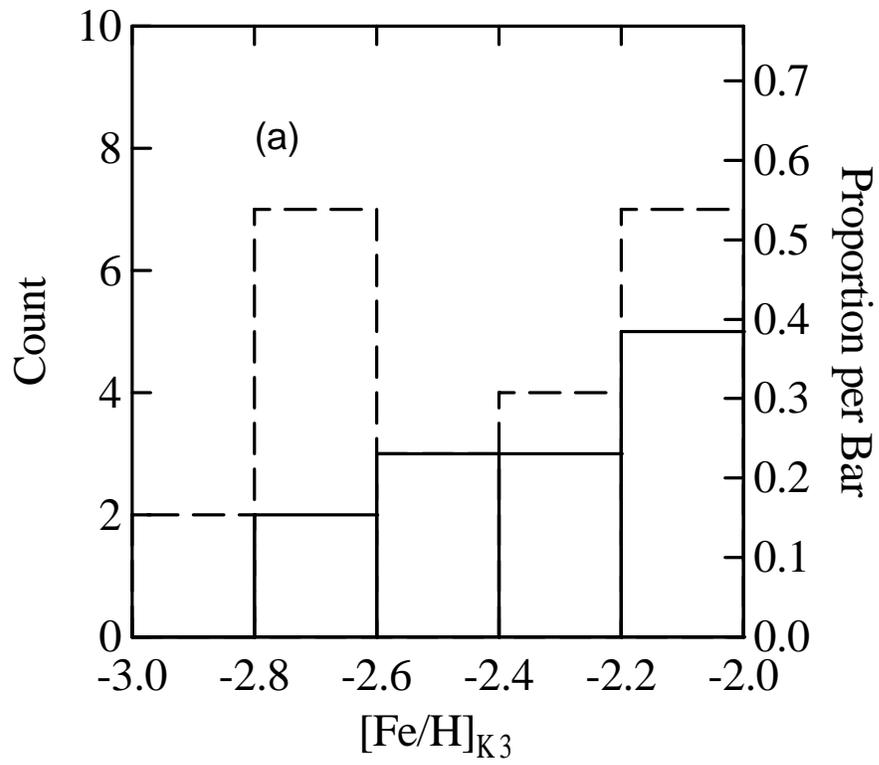
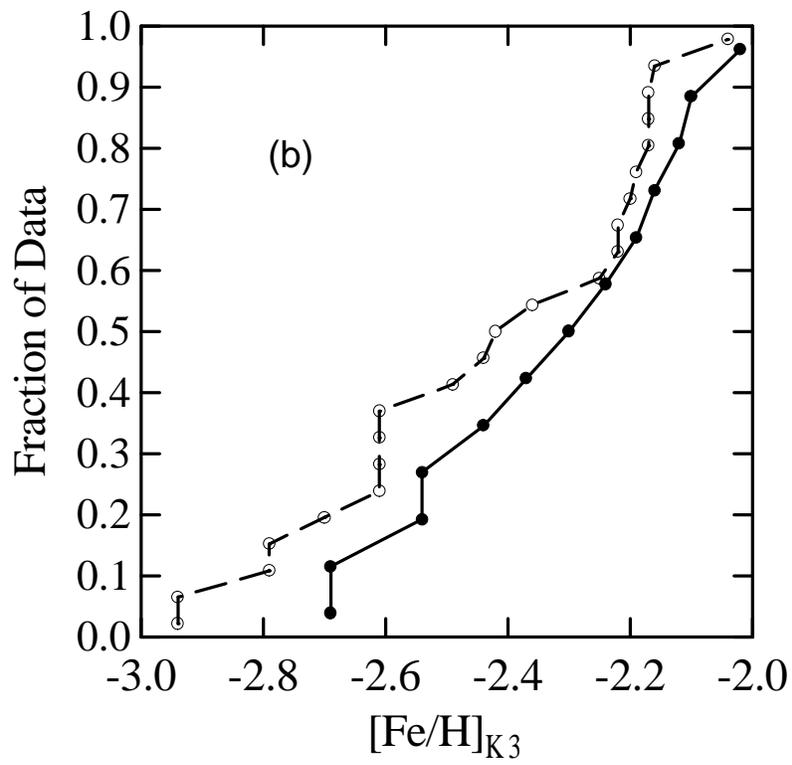
\end{document}